# Recent Progress in the Concurrent Atomistic-Continuum (CAC) Method and Its Application in Phonon Transport

Xiang Chen[1,*], Weixuan Li[1], Adrian Diaz[1], Yang Li[1], David L. McDowell[2,3] and Youping Chen[1]

[1]Department of Mechanical and Aerospace Engineering, University of Florida, Gainesville, FL 32611
[2]Woodruff School of Mechanical Engineering, Georgia Institute of Technology, Atlanta, GA 30332, USA
[3]School of Materials Science and Engineering, Georgia Institute of Technology, Atlanta, GA 30332, USA

**Abstract**

This work presents recent the progress in the development of the Concurrent Atomistic-Continuum (CAC) method for coarse-grained space- and time-resolved atomistic simulations of phonon transport. Application examples, including heat pulses propagating across grain boundaries and phase interfaces, as well as the interactions between phonons and moving dislocations, are provided to demonstrate the capabilities of CAC. The simulation results provide visual evidence and reveal the underlying physics of a variety of phenomena including: phonon focusing, wave interference, dislocation drag, interfacial Kapitza resistance caused by quasi-ballistic phonon transport, etc. A new method to quantify fluxes in transient transport processes is also introduced.

## 1. Introduction

Microstructure hierarchy and its dynamic behavior have increased the essential complexity of materials. However, there has historically been a lack of predictive computational tools in modeling, understanding, and ultimately controlling materials that exhibit complex mesoscale dynamic behavior. Simulating such materials requires bridging length scales from atomistic structures and their interactions to continuum models of bulk components[1]. The majorities of existing methods treat the mechanical and thermal behavior of materials separately and require empirical parameters from theoretical modeling or fits to experiments. Specifically, when dealing with heat conduction, assumptions are usually made regarding the nature of phonon thermal transport process, e.g., Fourier or non-Fourier heat conduction, ballistic or diffusive phonon transport, or a preconceived understanding of the scattering between phonons and defects; these assumptions limit the predictive power of these methodologies.

Among the existing computational tools for phonon transport, there are several widely used ones from different perspectives, which have yielded useful insight and have significantly enhanced our understanding. The atomistic Green's function method (AGF) pioneered by Mingo and Yang[2] has emerged for a decade as a useful tool for the study of phonon transport across interfaces. Similar to the lattice dynamics (LD) method, the AGF technique is a frequency domain method ideally suited for phonon mode analysis and frequency specific transmission across interfaces. It is based on the harmonic force constants from empirical or DFT calculations; the inclusion of anharmonicity, although possible in principle, requires significantly more effort and is not generally pursued in the literature[3]. AGF and LD can be applied to study the effects of defects but with the defect structure being considered as a time invariant input. Evolution of interfacial structure or other defects is not considered. Another widely used method for the study of phonon thermal transport is the Boltzmann transport equation (BTE). Along with the recent developed efficient Monte Carlo (MC) algorithm, this tool has been extended for the study of mesoscale material systems[4-5]. However, in BTE the defects are considered through the extrinsic relaxation time of phonons, and hence the effects of different defects are simply combined via Mathiessen's rule. This means that the scattering mechanisms of multiple defects, as well as the intrinsic phonon-phonon scattering, are all assumed to be independent of each other. In addition, the BTE technique treat phonons as particles, this makes it inappropriate for the study of wave effects, i.e., wave interference[6].

To model phonon transport in materials with defects, Molecular Dynamics (MD) is essentially the only tool currently available that does not require assumptions regarding the nature of phonon transport or mechanisms for phonon-defect scattering[7-8]. Equilibrium MD with the Green-Kubo algorithm can be used to compute thermal conductivity and, if projected onto harmonic normal modes, it can be used for a modal analysis[9]. However, since it assumes one heat flux for the entire specimen, the method is only appropriate for homogeneous material system[10]. The nonequilibrium steady-state MD is analogous to the quasi- one dimensional steady-state experiment and is straightforward in terms of data analysis. One component of the thermal conductivity tensor can be obtained in each simulation by fitting the temperature profile to Fourier's law[11-13]. MD can also be used to simulate transient transport processes, such as heat pulse propagation, if the heat pulse is properly modeled. The major limitation of using MD for the study of phonon-defect interaction is the size limit. The maximum phonon wavelength that can be considered in the simulation is the length of the MD simulation cell. Thus, the effect of long wavelength phonons cannot be investigated. However, phonon transport in short-period superlattices, for example, has experimentally been shown to be dominated by long-wavelength phonons[14]. Even for single crystalline Si at room temperature, phonons with micron-sized wavelength contribute more than 40% of the thermal conductivity[15].

The CAC method is developed to meet the need for space- and time-resolved simulation of highly nonequilibrium transport processes, addressing mesoscale phenomena in hierarchical materials, beyond the length scale limits typical of MD. This is motivated by recent advances in conducting transient experiments with direct photoexcitation that have led to observations of many new phenomena, while improving the understanding of the underlying physics[16-23]. The objective of this work is to introduce recent applications of the CAC method in simulating phonon transport. The rest of the article is organized as follows. The CAC methodology for simulation of phonon transport is presented in Section 2, including verifications of the method in predicting phonon dynamics. In Section 3, we present three numerical examples to demonstrate the capabilities of CAC in simulating ballistic and diffusive phonon propagation, their interaction with defects, as well as the dynamic responses of the defects. In Section 4, we present atomistic formula for heat flux that can be used to compute heat flux in transient processes and inhomogeneous systems. This paper is concluded with a brief summary and discussion in Section 5.

## 2. The CAC method

### 2.1 A brief introduction

The concurrent atomistic-continuum (CAC) method is a coarse-grained atomistic method that reduces the degrees of freedom of an atomistic system by reformulating the continuum field representation of balance laws from the atomistic model[24]. We emphasize here its fully dynamic form. The theoretical foundation of the CAC method includes a coarse graining concept based on the two-level description of crystals in solid state physics and a unified atomistic-continuum formulation of the conservation laws based on the principles of classical statistical mechanics. The solid state physics and statistical mechanics considerations can be summarized as follows.

(1) Solid state physics describes the structure of all crystals in terms of a periodic lattice with a basis of atoms attached to each lattice point[25] (cf. Fig.1). As the size of the lattice increases, the mechanical response associated with the structure become continuous[26]. CAC makes use of the translational symmetry of the crystal lattice and the lattice-level continuous properties to reduce the degrees of freedom of the atomistic model of crystalline materials. In contrast to many existing multiscale methods that coarse grain the atomic-scale structure or displacements, CAC reduces the degrees of freedom by assuming continuous deformation of the lattice while retaining information regarding discontinuous atomistic structure and response within any given unit cell in the case of polyatomic unit cells.

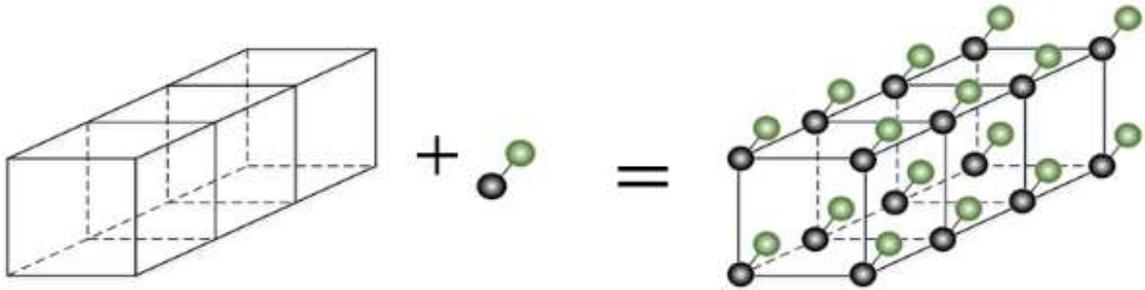

**FIG. 1 Solid state physics description: "crystal structure = lattice + basis" [27].**

(2) Statistical mechanics has been the major theoretical tool to formulate a coarse-grained description of the microscopic dynamics of particles. For nonequilibrium systems, most developments have followed from the work of Kirkwood on the "statistical mechanical theory of transport processes"[28-29]. In 1950, Irving and Kirkwood formulated the hydrodynamics balance equations based on a molecular model, in which a continuum field description of local densities, such as stress and heat flux, are linked to the mass, position, and momentum of molecules, with the internal degrees of freedom within a molecule being ignored. In the first of a series of 14 classic papers, Kirkwood envisioned the extension of his formulation to molecules possessing internal degrees of freedom[29]. The CAC formulation is such an extension to the description of crystalline materials. It describes a crystalline material as a continuous collection of lattice cells with a group of discrete atoms embedded within each lattice cell.

This two-level description is motivated by Kirkwood's idea of including the internal molecular degrees of freedom[29], the "lattice + basis" description of crystals in solid state physics[27], and also the two-level materials description in Micromorphic theory[30-35]. Following the Irving-Kirkwood procedure[28], this concurrent two-level description leads to a concurrent atomistic-continuum representation of the conservation laws[36-37] as exact consequences of Newton's second law[36-38], i.e.,

$$\frac{d\rho^\alpha}{dt}+\rho^\alpha(\nabla_x \cdot v+\nabla_{y^\alpha} \cdot \Delta v^\alpha)=0$$
$$\rho^\alpha \frac{d}{dt}(v+\Delta v^\alpha)=\nabla_x \cdot t^\alpha+\nabla_{y^\alpha} \cdot \tau^\alpha+f^\alpha_{ext} \quad (1)$$
$$\rho^\alpha \frac{de^\alpha}{dt}=\nabla_x \cdot q^\alpha+\nabla_{y^\alpha} \cdot j^\alpha+t^\alpha:\nabla_x(v+\Delta v^\alpha)+\tau^\alpha:\nabla_{y^\alpha}(v+\Delta v^\alpha)$$

where $x$ is the physical space coordinate of the continuously distributed lattice; $y^\alpha$ ($\alpha = 1,2,...,N_a$, where $N_a$ is the total number of atoms in a unit cell) is the subscale internal variable describing the position of atom $\alpha$ relative to the mass center of the lattice located at $x$; $\rho^\alpha$, $\rho^\alpha(v + \Delta v^\alpha)$, and $\rho^\alpha e^\alpha$ are the local densities of mass, linear momentum and total energy, respectively; $v + \Delta v^\alpha$ is the atomic-level velocity and $v$ is the velocity field; $f^\alpha_{ext}$ is the external force field; $t^\alpha$ and $q^\alpha$ are the momentum flux and heat flux due to the homogeneous deformation of lattice cells; $\tau^\alpha$ and $j^\alpha$ are the momentum flux and heat flux resulted from the reorganizations of atoms within the lattice cells. Solutions of the new conservation equations, supplemented by the underlying interatomic potential, provide both the continuous lattice deformation and the rearrangement of atoms within the lattice cells, thus leading to a concurrent atomistic-continuum methodology; in the two limiting cases, namely the atomic and the macroscopic scales, the atomistic and continuum descriptions of transport processes are recovered.

CAC is also a continuum field theory since it treats a crystalline material as a continuous collection of material points (unit cells) that possess internal degrees of freedom to describe the movement of atoms inside each unit cell. In distinction from the two-level continuum description in Micromorphic theory[30-35] or other generalized continuum theories[39-40], in CAC the subscale description within each continuously distributed material points is discrete, with atoms in each lattice cell being modeled explicitly. This complete set of governing equations governs both the atomistic and continuum regions, with the atomic-scale structure and interactions being built in the formulation. Different from the Irving and Kirkwood's statistical mechanics formulation of single phase, single component hydrodynamics equations, CAC employs a concurrent two-level structural description of materials. Therefore, CAC is unique in a number of regards:

(1) With a single precisely formulated set of conservation equations governing both atomic and continuum domains, the usual artificial interface between atomistic-continuum descriptions that limits most multiscale methods to consider only static phenomena is not needed;

(2) By including the internal degrees of freedom of atoms within each lattice cell in the continuum formulation, CAC can simulate complex crystalline materials, including both acoustic and optical phonon branches;

(3) With a complete field representation of the governing laws, the CAC formulation can be solved efficiently using continuum field simulation approaches, e.g., the finite element method. With the only constitutive relation being the interatomic potential and with a recast form of the governing equations, continuity between elements is not required; consequently, nucleation and propagation of dislocations or cracks can be simulated via sliding and separation between finite elements, as direct consequences of the governing equations.

Since CAC is derived solely from an atomistic model, it does not require continuum parameters for static, dynamic, zero or finite temperature systems. In the process of the derivation, the fluxes are determined in terms of atomic displacements and local temperature. The CAC formulation is currently numerically implemented using a modified finite element method. It differs from traditional finite element implementations of classical continuum mechanics since each finite element in CAC represents a collection of continuously deformable primitive unit cells, each containing a group of atoms. Applications of CAC for the simulation of mechanical behavior have been demonstrated through the reproduction of dynamic phenomena such as crack propagation and branching[41-43], phase transitions[44], nucleation of dislocations, formation of dislocation loops and networks[45-51], and defect-interface interactions[52].

## 2.2 Extending the form and capability of CAC to phonon transport

A recent development of CAC is to model coupled phonon dynamics and defect dynamics for nonequilibrium transient processes in heterogeneous materials. Note that CAC can reproduce the exact dynamics of an atomistic system if modeled with the finest mesh. When discretized with coarse-scale finite elements, CAC can accurately predict the dynamics of phonons with wavelengths longer than a critical wavelength determined by the size of the element.

To enable CAC to simulate heat pulse experiments, we have developed a phonon representation of heat pulses that are composed of spatiotemporal Gaussian wave packets, termed the coherent phonon pulse (CPP) model[53]. The CPP model is used to mimic the coherent excitation of a nonequilibrium phonon population achieved via ultrashort laser techniques. Since the CAC method is a coarse-grained method derived bottom-up from atomistics, and can adapt the viewpoint of phonons to describe heat and temperature in terms of mechanical vibrations, the CPP model naturally accommodate to the framework of CAC. In addition, since CAC is built on the statistical mechanical theory of transport processes of Irving-Kirkwood[28-29], it can naturally be used to study highly nonequilibrium dynamic phonon transport processes.

The current CAC method can be used for coarse grained study of long wavelength phonon dynamics and the dynamic interaction between phonons and defects. The propagation of heat pulses or defects, including their propagation directions and speeds, all emerge naturally in the simulation without the need for ballistic diffusive transport assumptions, or *a priori* understanding of the mechanism of phonon-phonon and phonon-defect interactions.

## 2.3 Verification of CAC in simulating phonon transport

### A. Phonon dispersion relations

For the purpose of verification, we perform the LD calculation of the phonon dispersion relations for a 2D FCC solid with a Lennard-Jones (LJ) interatomic force field[54] and compare the results to the dispersions in CAC obtained by solving the dynamical matrix of the coarse-grained lattices. Figure 2 presents the frequency-wavelength relations along the [110] direction of the FCC crystal, containing both longitudinal acoustic (LA) and transverse acoustic (TA) branches. It is seen that the phonon dynamics predicted by CAC match the LD solution well within the range of the allowable wavelength. Here, the allowable wavelength (10 nm to 1 μm) is determined by the sizes of the specimen and the finite element mesh with an allowed error (<5%). The characteristic length of the specimen is ~1 μm, and the finite element contains 8 unit cells along each direction.

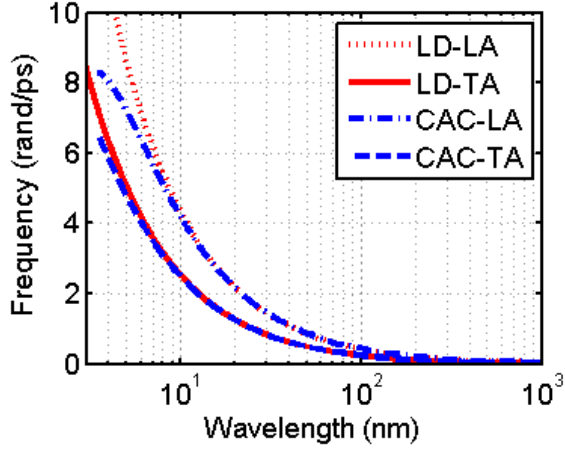

**FIG. 2 Phonon dispersion relations along the [110] direction within the (111) plane of a FCC single crystal. The solutions calculated for the coarse-grained lattices in CAC (in blue) are compared to the lattice dynamics solution (in red).**

## B. Monochromatic wave propagation

Motivated by the experimental observations of radial phonon distributions in materials induced by the photoexcitation[55], most of the numerical examples presented in this work are based on a point heat source propagation in a 2D computer model. In this section, we verify the simulation tool by considering propagation of a continuous monochromatic radial wave (20 nm in wavelength) in a 2D model, and comparing the simulation results with MD. As can be seen in Fig. 3, when the mesh size approaches full atomic resolution, the CAC simulation with fine mesh generates the same results as MD. Note that the limited specimen size in the middle and for right columns induces surface reflection. With coarse-graining, a larger computer model is simulated, and the exploded view of the same region around the heat source is presented in the left column for comparison. It is seen that the phonon wavelength, group velocity, and the slight hexagonal shape of wave crests caused by phonon focusing are reproduced by CAC with a coarse mesh in agreement with MD results.

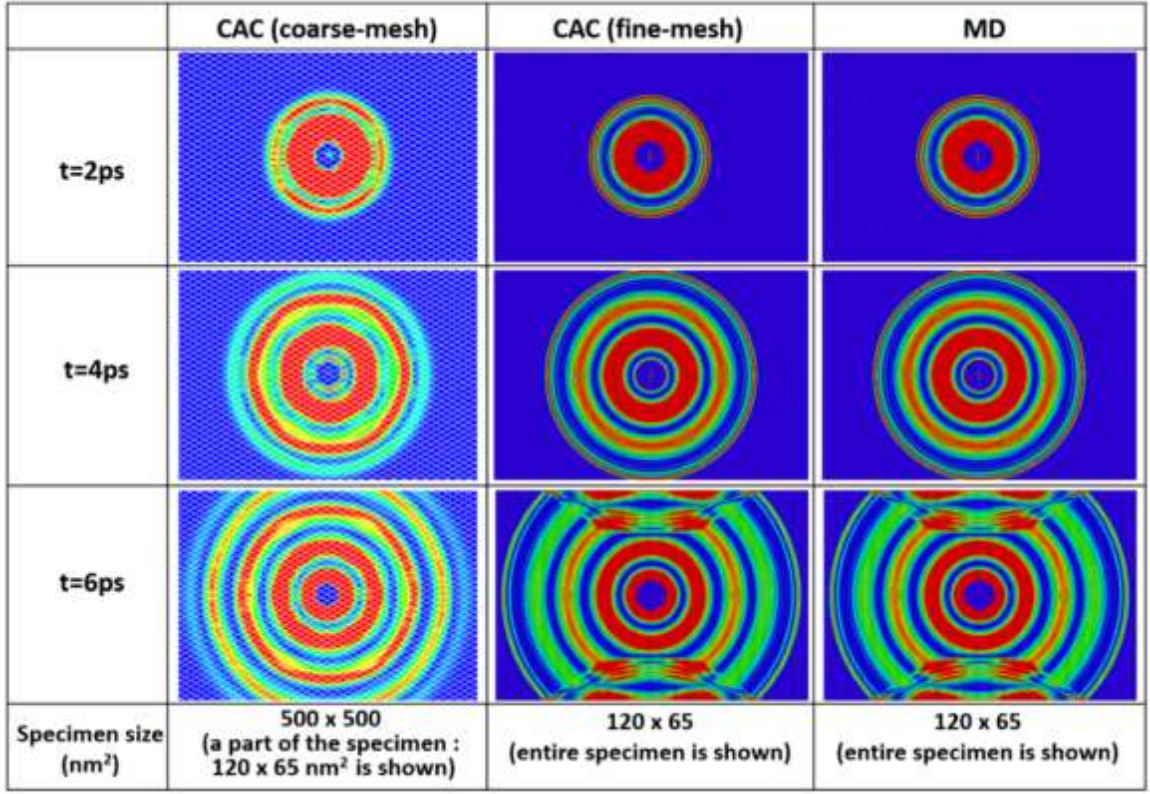

**FIG. 3 Comparison of a monochromatic wave propagation in CAC with coarse-mesh, fines-mesh and in MD.**

## C. Passing long wavelength phonons through the numerical interfaces

To verify that there is no spurious wave reflection at the atomistic-continuum interface for the long wavelength phonon propagation, the single crystal model is discretized with both uniform meshes and non-uniform meshes, i.e., with both coarse meshes and fully atomistically-resolved meshes. The CAC simulation is then conducted along with a CPP model[53], which continuously generate phonons with wavelengths ranging from 5 nm to 250 nm. The heat source is applied to the center of the models where the coarse elements are employed; the spot size of the heater is 10 nm to avoid strong phonon-phonon scattering.

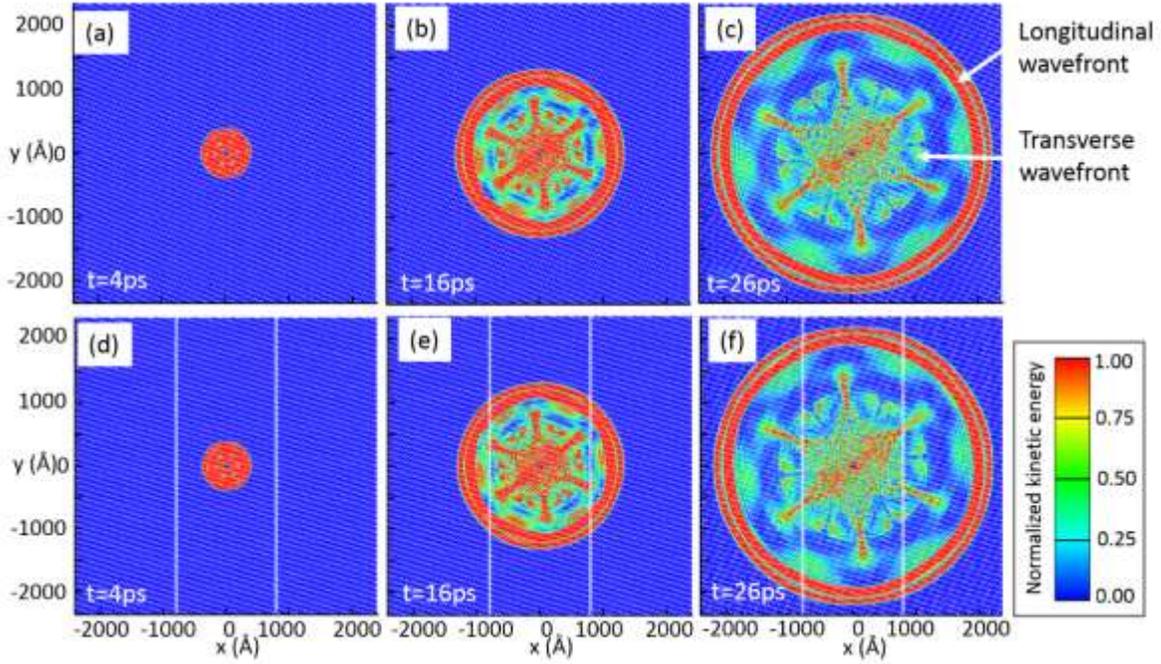

**FIG. 4 Comparison of the phonon wave propagation in a single crystal specimen constructed with uniform coarse meshes (a-c) and non-uniform meshes with both atomic and coarse-scale finite element meshes (d-f).**

In Fig. 4, the normalized kinetic energy distributions in the single crystal models with uniform meshes (Fig. 4(a-c)) and non-uniform meshes (Fig. 4(d-f)) at the same time steps are compared; there is no noticeable difference. The maximum error is then quantified by comparing the pointwise total kinetic energy in different regions, which is less than 0.5%. This result confirms that the atomistic-continuum interface in CAC does not provide a numerical barrier for the dynamic propagation of long wavelength phonons. It is also noticed that relatively short wavelength phonons channel along the [112] crystal directions and result in phonon-focusing "caustics", which are especially high phonon fluxes along certain directions arising from crystallographic anisotropy. This indicates that CAC method is able to capture experimentally observed phonon focusing phenomena[55]. In addition, the longitudinal and transverse wavefronts are indicated in Fig. 2(c). The associated wave speeds are measured from the time sequence, as ~6700 m/s and ~3900 m/s, respectively, which are consistent with the phonon group velocities of the LA and TA phonons near the $\Gamma$-point from LD calculations. These verifications once again demonstrate the capability of CAC to predict long wavelength phonon dynamics in crystalline materials.

3. **CAC simulations of coupled phonon transport and defect dynamics**

## A. phonon transport across grain boundaries

Interfaces play a significant role in thermal transport in crystalline materials. However, modeling and simulation of phonon-interface interaction has been historically challenging, especially for incoherent interfaces such as high-angle grain boundaries (GBs)[56]. Moreover, coupling transport simulation to structural simulation has been identified as a significant challenge in phonon transport[6]. In this section, we demonstrate the capabilities of CAC to simulate the dynamic interaction of a heat pulse with GBs.

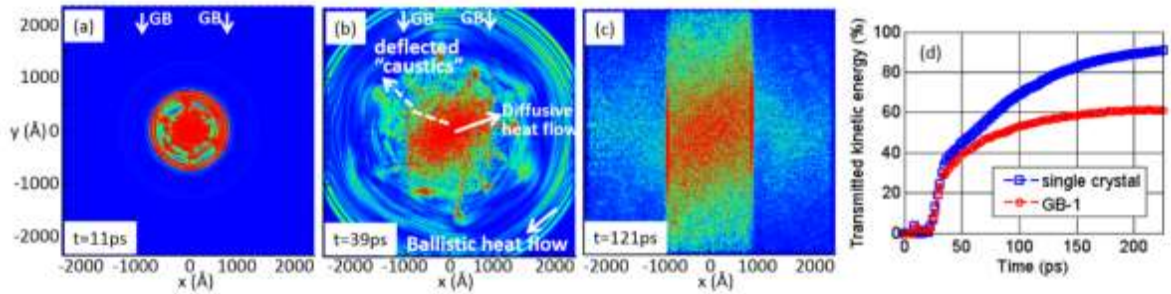

**FIG. 5 Transient phonon heat propagation in the GB-1 model. (a)-(c) present the time sequences of the kinetic energy distribution in the GB-1 model. The transmitted kinetic energy of the GB-1 mode and that of the single crystal model during the dynamic process are presented in (d).**

Figure 5 (a-c) present the time sequences of the normalized kinetic energy obtained from CAC simulations of the propagation of a heat pulse across GBs, with the inclination angle between the crystalline grains being 13.2°. More details regarding the computer models and the simulation results have been reported in one of our recent works [57]. Figure 5 clearly demonstrate the coexistence of ballistic and diffusive phonon transport, as well as the coexistence of coherent and incoherent phonon scattering. Figure 5(d) presents the energy transmission across the specimen with and without GBs. This comparison, along with the discontinuities in kinetic energy shown in Fig. 5(c), provides a quantitative understanding of the GB resistance to heat flow. The combination of the diffusive scattering of coherent and incoherent phonons and the reflection of the coherent phonons by the GBs leads to the Kapitza resistance to heat flow in the simulated specimen, as shown in Fig. 5(a-c). Moreover, with a selected input phonon mode, the wavelength-dependent energy transmission has been measured [57], which shows the capability of CAC to conduct a mode-wise study.

In addition to the qualitative and quantitative study of the effect of GBs on the phonon transport, the spatial and temporal resolved simulation results from CAC can simultaneously capture the defect structural evolution. As shown in Fig. 6(a-d), during the dynamic process, we find that GBs undergo local reconstruction when meeting with phonon caustics of high energy flux. Depending on the phonon intensity, the reconstruction may occur with or without recovery. The four figures on the top of Fig. 6(a-d) are the exploded views, showing the atomic details of a local GB region during the dynamic interaction with phonons. Note that the white dashed lines mark the initial position of the GB. These simulation results demonstrate that the defect structures can dynamically respond during the phonon transport process. The CAC method along with the CPP model is thus demonstrated to provide a fully coupled treatment of defect dynamics and phonon transport; this is nontrivial for other existing methods that either treat a defect as a static input or consider it through the extrinsic scattering rate of phonons. Also, Fig. 6 points to the significance of long wavelength phonons on the dynamic evolution of microstructures.

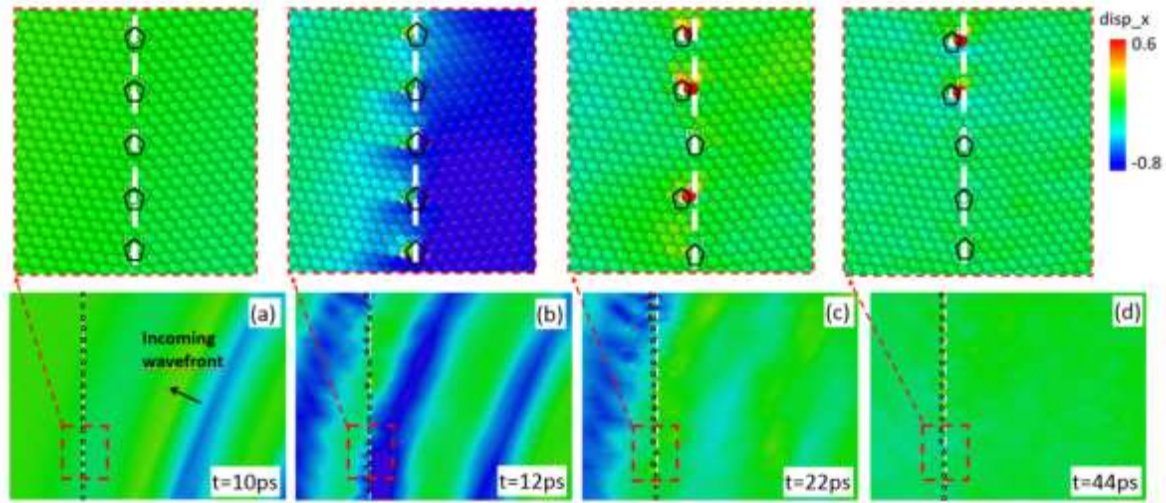

**FIG. 6 Time sequences of atomic arrangements showing the GB structural change due to the phonon-GB scattering; (a-d) are snapshots taken at the vicinity of the GB region where the GB intersects with the energy caustics of the phonon pulse. Corresponding to each figure, there is an exploded view of the highlighted region (in red dash line). The white dashed line marks the initial position of the GB.**

In addition, simulation results show that the majority of the heat is carried by long wavelength phonons that were not believed to be the major heat carrier. Distinct from the heat propagation in a single crystal, in a polycrystalline material, most of the short wavelength phonons far away from the $\Gamma$-point are confined within the heat-source-grain by the GBs. Therefore, it is the propagation of long wavelength phonons across the GBs outside the heat-source-grain that determines the capability of a polycrystalline material to conduct heat via phonons. The simulation results demonstrate the capability of CAC in the simulation of heat pulse propagation in polycrystalline materials via phonon transport.

## B. Wave interference in superlattices

Wave interference is believed to play a significant role in phonon transport in periodic superlattices. However, detecting this transient process requires extremely fine temporal and spatial resolutions; a direct experimental observation is thus prohibited. Instead, indirect supporting evidence, such as the existence of the minimum thermal conductivity as a function of the superlattice period length[58-63] and the linear size-dependence of the superlattice thermal conductivity[64], are considered to be indicative of the phonon wave interference effect.

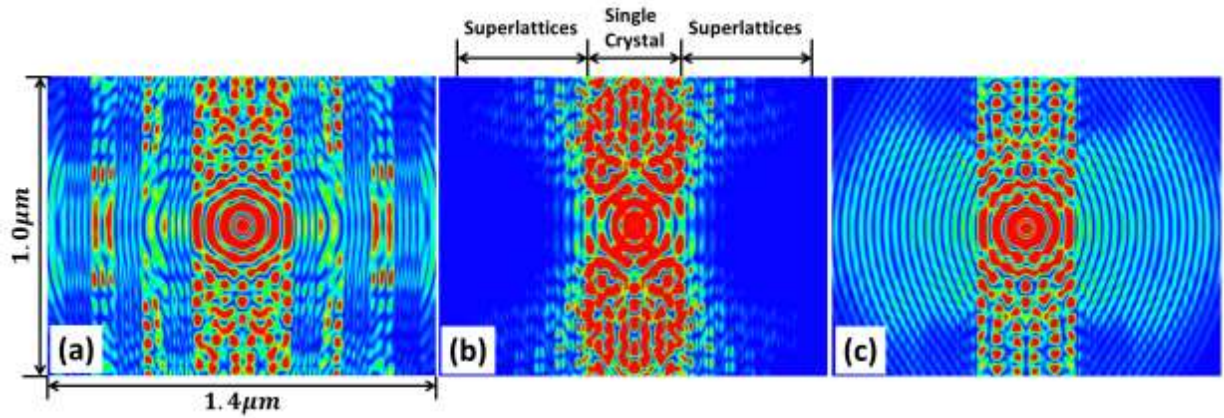

**FIG. 7 Snapshots of normalized kinetic energy distribution during the propagation of a same monochromatic phonon wave (with a wavelength of 106.3 nm) in three superlattices of equal size but different period length (a) 177.1 nm, (b) 35.4 nm, and (c) 17.7 nm.**

In this section, we demonstrate the capability of CAC to provide direct visualization of the wave interference phenomena in superlattices. Computer models of superlattices are constructed with two types of materials (A and B) governed by LJ interatomic potential. The ratios of the atomic masses, the zero-crossing distances and the dielectric constants for the two materials are respectively $m_A/m_B = 1.2$, $\sigma_A/\sigma_B = 1.0$ and $\varepsilon_A/\varepsilon_B = 5.0$. Three superlattices of equal size but different period length are modeled and simulated. With the propagation of a monochromatic phonon wave (106.3 nm in wavelength), the kinetic energy distribution in the three models presented in Fig. 7(a-c) show distinct responses. In Fig. 7(a), the phase interfaces in the superlattice structure behave independently and the wavelength changes alternatively in different material layers. In Fig. 7(b), we observe that the phonon wave propagating in the direction perpendicular to interfaces is completely reflected back by the interfaces; this provides a direct visual evidence of Bragg reflection[65]. Figure 7(c) shows that the superlattice behaves in this case like a homogeneous material, with the phonon wave traveling across it ballistically. These simulation results demonstrate that, without an assumption regarding the nature of phonon transport, i.e., whether it is particle-like or wave-like, the wave-like transport phenomenon naturally emerges in the CAC simulation of superlattices.

## C. Phonons scattering by moving dislocations

Simulation of fast moving dislocations is nontrivial. Most of the existing studies have considered dislocations under static or quasi-static conditions[66-72]. For the study of multiple dislocations, the dislocation spacing is usually assumed to be sufficiently large so that the interactions are decoupled[73]. To provide understanding of the fast moving dislocation at the atomic level, MD has demonstrated its predictive capability. However, the long-range nature associated with the phenomenon of moving dislocations, especially multiple moving dislocations, as well as their interaction with propagating phonons, is beyond the length scale of MD at reasonably affordable computational cost. The CAC method can be used to complement MD to provide this information. A verification study of the CAC method in modeling fast moving dislocations has been reported[74]. One of our recent works has presented a phonon pulse propagation in crystalline materials with a fast moving dislocation and dislocation array while studying the effect of phonons on dislocation mobility[75].

Figure 8 presents the kinetic energy distributions at the vicinity of the dislocation cores during the process of nucleation and migration of a single dislocation (Fig. 8(a-d)) and an array of multiple dislocations (Fig. 8(e-h)). The time sequences show that the dislocation motion is accompanied by a V-shape pattern of strong lattice vibrations in the wake of the moving dislocation, together with the relatively weak radial-shape wave front; this is consistent with a recent MD study on a single dislocation[76]. The dynamic process presented in Fig. 8 also shows that the radial wave front propagates faster than the dislocation core, indicating subsonic dislocation motion. This also explains why the wave interference band in Fig. 8(f-h) widens as the dislocation array advances. In addition, our simulation results show that a dislocation can travel to microns before its velocity reaches a steady state condition[75]; this explains the size-effect in MD[77-78].

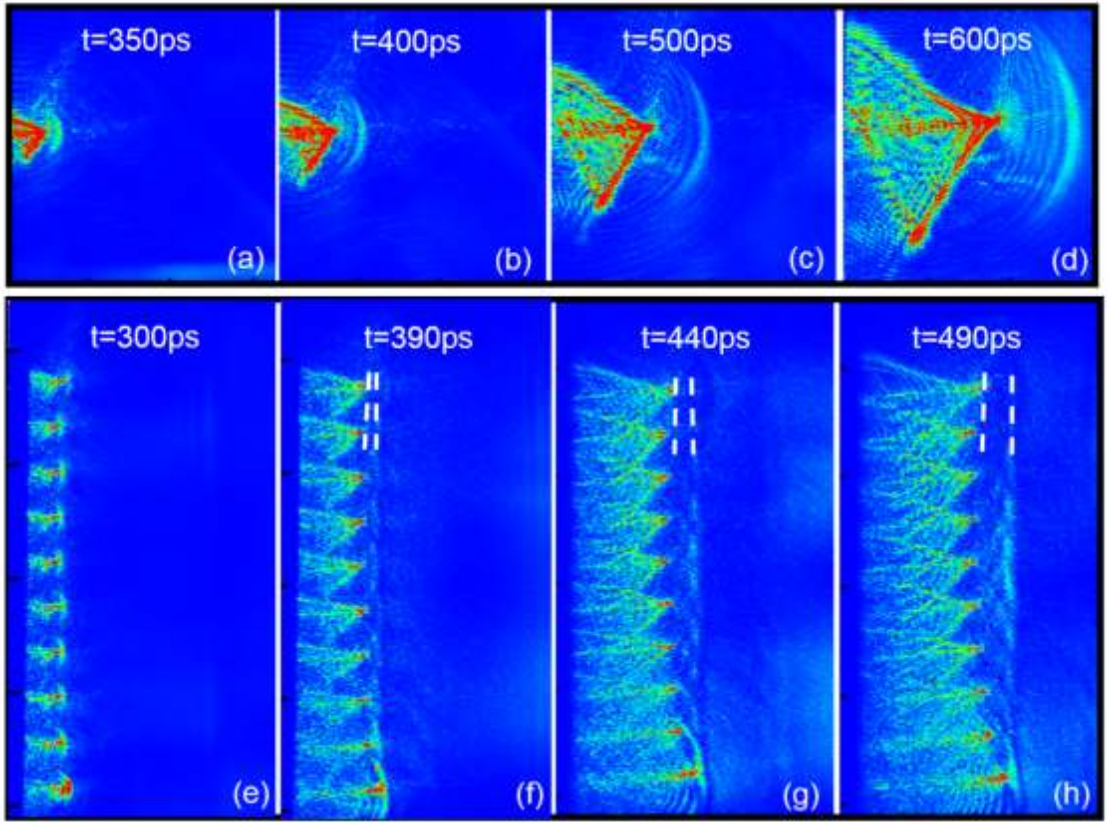

**FIG. 8 Time sequences of the kinetic energy distributions at the vicinity of the dislocation cores during the process of nucleation and migration of a single (a-d) and multiple dislocations (e-h), and phonon waves emitted from dislocations.**

Figure 9 presents the interaction between a moving dislocation array and a propagating phonon pulse. The domain size is 5 µm by 5 µm. It is shown that during the process, multiple phonon scattering mechanisms coexist and cannot be simply decoupled. Therefore, it is important to naturally incorporate all the scattering events without additional assumptions regarding either the individual scattering mechanism or a rule that combines them. For example, with multiple dislocations, the strong wave interference band scatters the incoming phonon pulse before the phonons meet with the dislocations, as shown in Fig. 9(a). This in turn, changes the mobility of the dislocation when it is in the array compared to when it is present alone[75]. Figure 9(b) shows that during the interaction, one of the dislocations starts to decelerate by emitting phonon waves from the dislocation core and is soon arrested. After the wave front passes by, as shown in Fig. 9(c), three dislocations (with red dislocation marks) out of ten are arrested. The simulation of fast moving dislocations interacting with the phonon pulse once again demonstrates that CAC is able to provide a fully coupled treatment of defect dynamics and phonon transport. In addition, the spatial and temporal resolved simulation results have allowed a detailed investigation of the dynamic processes and their underlying physics.

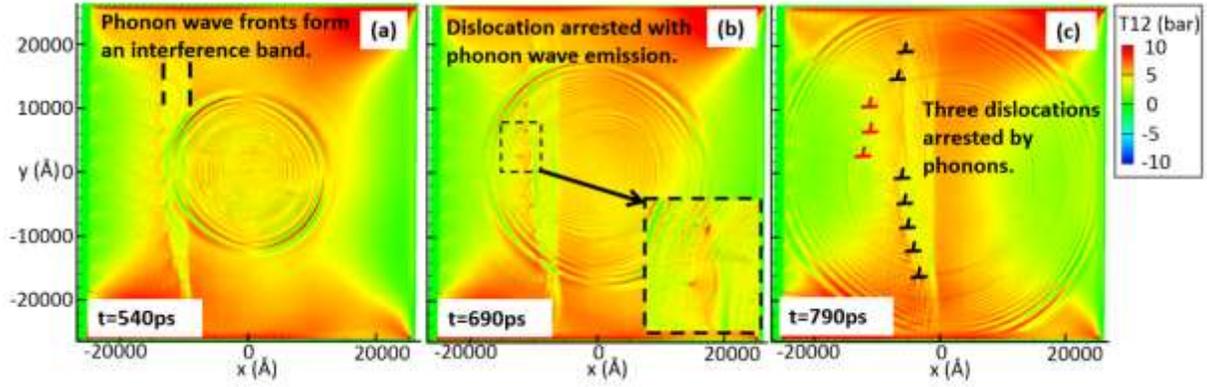

**FIG. 9** Time sequences of the shear stress distributions show the propagating heat pulse interactions with multiple moving dislocations. (a) Multiple dislocations moving in an array and a phonon pulse propagates towards the moving dislocations. (b) After interaction with the phonon pulse, one of the dislocations starts to decelerate through wave emission. (c) Three out of ten moving dislocations are arrested after interaction with the phonon pulse.

### D. Heat flux formula for transient transport processes and inhomogeneous materials

Momentum and heat fluxes in the majority of literature and in popular MD simulators are defined as "a quantity per unit volume". The most popular expressions of such volume-averaged fluxes include the virial theorem for stress, the heat theorem for energy flux[79-80], and many atomistic formulations that followed the Irving-Kirkwood's statistical mechanical formalism[28] such as the Hardy formulas[81]. However, defining a flux as a quantity per volume is a fundamental departure from the physical definition of a flux as "a physical property across a surface per unit area and time"[28]. Consequently, these volume-averaged flux formulas have been shown to be inapplicable to transient transport processes or highly inhomogeneous systems[10, 82].

CAC is fundamentally a space and time resolved simulation tool. It aims for simulations of non-equilibrium transient processes in mesoscale hierarchical materials. This requires that the methodology for measuring fluxes be free of constitutive relations, except for the interatomic potential, and that space and time-resolved fluxes in the transient nonequilibrium simulations can be accurately measured. For this purpose, a new method is recently formulated for formally deriving microscopic momentum and heat fluxes through the integral conservation laws[10]. The resulting flux formulas are defined as surface averages, i.e.,

$$\sigma(x,t) = -\sum_k m_k \tilde{v}_k^\alpha \tilde{v}_k^\beta e^\alpha \bar{\delta}_{AT}^\beta(r_k-x) - \sum_{k,l} \frac{\partial \Phi_l}{\partial r_k^\alpha} \int_{L_{kl}} \bar{\delta}_{AT}^\beta(\varphi-x)d\varphi\, e^\alpha e^\beta \,, \qquad (2)$$

$$q(x,t) = -\sum_k \tilde{v}_k^\alpha [\tfrac{1}{2} m^k (\tilde{v}_k)^2 + \Phi_k] \bar{\delta}_{AT}^\alpha(r_k-x) + \sum_{k,l(\neq k)} \frac{\partial \Phi_l}{\partial r_k} \cdot \tilde{v}_k \int_{L_{kl}} \bar{\delta}_{AT}^\alpha(\varphi-x)d\varphi\, e^\alpha \,, \qquad (3)$$

where $m_k$, $r_k$, $v_k$, $\Phi_k$ are the mass, position, velocity, potential energy of the $k$-th particle, respectively; $e^\alpha$ or $e^\beta$ denotes the unit basis, and $\alpha$ and $\beta$ are summation indices; $\bar{\delta}_{AT}^\alpha(r_k-x)$ is the averaged Dirac Delta over time-interval $T$ and surface element $A^\alpha$ whose normal is $e^\alpha$. The quantity $\int_{L_{kl}} \bar{\delta}_A^\alpha(\varphi-x)d\varphi$ is the line integral of the averaged Dirac Delta over the line segment $L_{kl}$ that links particle $k$ and $l$; it is typically equal to one or zero based on whether the pair of particles intersects the surface.

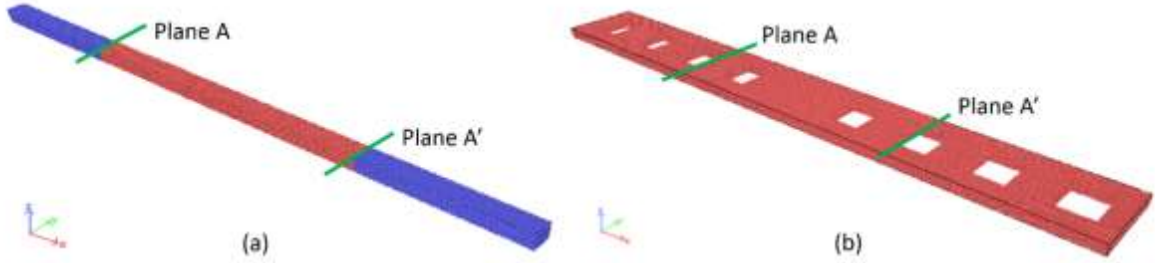

**FIG. 10 Computer model of (a) a superlattice structure with length 100 nm and a cross section of 3x3 nm; (b) a symmetric Si structure of dimension 2400x270x50 Å with holes. The interfaces at which planes intersect the model cross section to measure flux are denoted by the green lines.**

These new surface-averaged flux formulas have been shown to reproduce uniquely known values of the stress and heat flux in the equilibrium state, the steady-state, and transient transport processes. Here we provide simulations of two inhomogeneous systems to demonstrate the applicability of the surface-averaged formula for energy flux in transient transport processes: a Lennard-Jones superlattice structure in Fig. 10(a) and a Stillinger-Weber Si phononic crystal in Fig. 10(b).

The results of energy flux from different formulas are plotted in Fig. 11 and Fig. 12. The results using the surface-averaged energy flux formula, labelled as "*surface*", are compared with the energy flux that satisfies the Continuity equation, labelled as "*continuum*", and the volume-averaged heat flux formula, labelled as "*volume*". Here, the "*continuum*" energy flux formula is based on the hydrodynamics continuity equation through measuring the change of the total energy in a region to obtain the flow of energy across the entire surface enclosing the region[83], it is also the formula widely used in nonequilibrium MD simulations with the heat-source-sink scheme to calculate the heat flux in steady-state heat flow. Meanwhile, the "*volume*" formula is widely used in equilibrium MD simulations using the Green-Kubo formalism to compute thermal conductivity. Since the Hardy fluxes are volume averages that produce similar results to the virial and heat theorems[84, 85], we only plot the energy flux based on the heat theorem as the medium of comparison for volume averages. It is noted that the "*continuum*" formula measures the total energy flux across the entire surface rather than the local energy flux. However, with periodic boundary conditions being applied in both the *y* and *z* directions, the total energy flux averaged over two symmetrically located planes A and A' is equal to the averaged local energy flux over the area of the two planes.

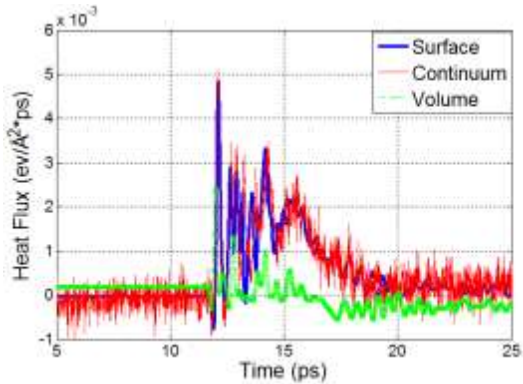

**FIG. 11 Energy flux during the transient simulation of the propagation of a heat pulse, averaged over two symmetric planes A and A'. Each data point is averaged over 100 time steps. The pulse inputs a total of 100 eV over 10 fs to the center region of the model.**

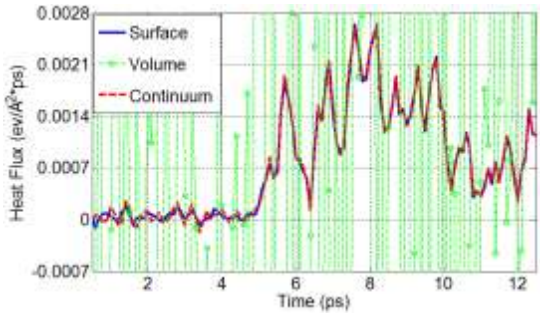

**FIG 12 Energy flux during the transient simulation of the propagation of a heat pulse, averaged over two symmetric planes A and A'. Each data point is averaged over 100 time steps. The pulse inputs a total of 100 eV over 10 fs to the center region of the model.**

Both the simulation results of the superlattice and the phononic crystal show that the transient energy flux across the interfaces predicted by the "*surface*" formula compares well with that of the "*continuum*" formula during the propagation of the heat pulses. By contrast, the "*volume*" formula significantly underestimates the transient energy flux in the superlattice, as shown in Fig. 11, and exhibits spurious fluctuations in the phononic crystal model, as shown in Fig. 12, when attempting a transient local computation of heat flux. These simulations demonstrate that the surface formula is in accordance with the total energy rate of change experienced by the chosen volume, and is capable of measuring energy flux in space- and time- resolved atomistic or coarse-grained atomistic simulations.

## 4. Summary and discussions

In summary, we have presented recent progress in the development and application of the Concurrent Atomistic-Continuum (CAC) method for modeling phonon transport. The CAC method's predictive capabilities in the realm of phonon dynamics are verified through (1) comparison of phonon dispersion relations obtained by CAC with that by lattice dynamics, (2) agreement of wave propagation in CAC with that in MD, and (3) phonon transport across the numerical interfaces between fully resolved atomistic and coarse-grained continuum domain. The CAC method is shown to be able to accurately reproduce the dynamics of an atomistic system if modeled with the finest mesh and, if discretized with coarse-scale finite elements, it can predict the dynamics of phonons with wavelengths longer than a critical wavelength determined by the size of the element. The atomistic-continuum interface does not act as a numerical barrier or induce spurious wave reflection for long wavelength phonons. Three numerical examples are then presented to demonstrate the capability of CAC, a few challenges in the modeling of phonon heat transport in materials with microstructure are addressed:

1) Since CAC is derived bottom-up from atomistics in phase space, the structure of defects can naturally be resolved in the real space. This is critically important for the study of defects with irregular structures and large characteristic length scale, such as incoherent grain boundaries and phase interfaces, multiple dislocations, etc.

2) As CAC naturally couples structural dynamics to phonon dynamics, it enables simulation of the dynamic evolution of defects in materials, such as the grain boundary reconstruction and dislocation drag, in transient processes of phonon transport.

3) In CAC, if multiple defects are modeled in one specimen, all scattering events are present in one simulation; coupled scattering mechanisms can thus be simulated without any assumption such as the Mathiessen's rule employed in the BTE method.

4) The physical space- and time- resolved simulation results provide visual evidence for a variety of phenomena without any *a priori* understanding or assumptions regarding phonon transport or phonon-defect scattering. All the dynamic processes involved in the CAC simulation are determined only by the Newton's law of motion, the interatomic potential, and the material microstructures.

5) With the input of a selected phonon mode, CAC simulation allows a mode-wise study of the microscopic details which can answer the fundamental questions such as which phonon-phonon or phonon-defect interactions are primarily responsible for a certain material property; e.g., a low thermal conductivity, how energy is distributed among the broad spectrum of phonons, etc.

Quantum-mechanical computational packages based on the Density Function Theory (DFT) allow a predictive, parameter-free study of the role of atomic bonding and phonon-phonon scattering[86], which can be used to predict the thermal conductivity of simple crystals. Although the extension to complex crystals with a rigorous treatment of disorder and boundaries requires further advances, accurate interatomic potentials DFT may provide can further enhance the predictive power of the atomistic simulations, as well as the coarse-grained atomistic simulations like CAC. For defects with small characteristic length, simulation tools such as MD can simultaneously capture the full atomistic picture of both the phonon thermal transport and structural evolution. To capture that picture for defects and structures with larger characteristic length, e.g., interfaces or moving dislocations, or materials with multiple defects, a coarse-graining technique such as that used in CAC holds the promise to provide useful insights.

Recent advances in experimental and computational techniques have enabled an unprecedented microscopic view of materials mechanical and thermal transport. However, many aspects of the fundamental microscopic details, as well as the connection between experimental observations and the simulation results, are lacking[5, 87]. Collaboration among these communities is needed to integrate information from investigations performed at different scales and/or from different perspectives. CAC aims to meet the need for a simulation tool that fully couples material microstructure, mechanical, and thermal transport behavior, and is capable of including both coherent and incoherent scattering processes in materials with microstructure, as well as defect dynamics and phonon transport.

**Acknowledgements**

This material is based upon research supported by the U.S. Department of Energy, Office of Science, Basic Energy Sciences, Division of Materials Sciences and Engineering under Award # DE-SC0006539.